Inelastic Scattering and Spin Polarization in Dilute Magnetic Semiconductor (Ga,Mn)Sb

Raghava P. Panguluri, B. Nadgorny

Department of Physics and Astronomy, Wayne State University, Detroit, MI 48201

T. Wojtowicz

Institute of Physics, Polish Academy of Sciences, 02-668 Warsaw, Poland

X. Liu, J. K. Furdyna

Department of Physics, University of Notre Dame, Notre Dame, IN 46556

ABSTRACT

The Point Contact Andreev Reflection (PCAR) technique has already been used to measure the spin polarization of some of the dilute magnetic semiconductors, such as narrow-band (In,Mn)Sb, as well as wider gap (Ga,Mn)As. While in (In,Mn)Sb conventional Andreev reflection has been demonstrated, in (Ga,Mn)As quasiparticle density of states (DOS) broadening has been observed, possibly due to inelastic scattering effects. Here, we investigate the spin polarization, magnetic, and transport properties of epitaxially grown (Ga,Mn)Sb films with the Curie temperature of ~ 10K. The spin polarization of 57±5% was measured. Spectrum broadening in (Ga,Mn)Sb has also been observed.

The electronic spin rather than charge can bring to life many fundamentally different devices, going beyond conventional semiconductor-based electronics.[1,2] While it is possible to use metals as spin injectors and detectors, the building blocks of these devices, it is much more natural, even from a purely fabrication standpoint, to use all-semiconductor systems. This is one of the main reasons dilute magnetic semiconductor (DMS) III-Mn-V compounds[3] have emerged as one of the most important classes of materials in semiconductor spintronics. (Ga,Mn)As[4,5] and related compounds are particularly well-matched with GaAs, traditionally used in electronic industry. However, while there have been extensive studies aimed at increasing their Curie temperature, $T_C$, it is still significantly lower than room temperature. In (Ga,Mn)N[6] $T_C$ has been predicted to exceed 300K,[7] but these DMS compounds typically have higher resistivity and thus are not well-suited as spin injectors.

The narrow gap materials, such as (In,Mn)Sb[8] and (Ga,Mn)Sb[9] are much less thoroughly investigated. In spite of their low $T_C$, these compounds have potential applications in spin dependent photonics, such as infrared or far-infrared devices, as well as in carrier-mediated spin dependent transport due to the presence of lighter, high mobility holes. As the effectiveness of a spintronic device is, to a large extent, dependent on the spin polarization $P$ of the spin injector,[10] it is important to determine the values of $P$ for this class of DMS materials. The spin polarization of (In,Mn)Sb epitaxial film has recently been measured[11] by the Point Contact Andreev Reflection (PCAR) technique.[12,13] In this letter we report the measurements of the spin polarization of (Ga,Mn)Sb thin films epitaxially grown on GaAs substrates, with $P$ evaluated to be 57±5%.

PCAR has recently been introduced as a technique to measure the transport spin polarization of various magnetic materials. In a normal metal/superconductor (N/S) junction, a quasiparticle with the energy less than the superconducting gap, $\Delta$, can convert into a Cooper pair and enter the superconductor by reflecting at the interface as a hole with (roughly) the same energy and opposite spin (Andreev reflection). In a ferromagnet (F) due to the spin-imbalance not every spin-up channel has the equivalent spin-down channel, leading to a partial suppression of Andreev reflection in the F/S junction, which can be detected from the conductance measurements. The spin polarization is given by $P_n = \dfrac{<N_\uparrow(E_f)v^n_{f\uparrow}> - <N_\downarrow(E_f)v^n_{f\downarrow}>}{<N_\uparrow(E_f)v^n_{f\uparrow}> + <N_\downarrow(E_f)v^n_{f\downarrow}>}$, with $n = 1$ in the ballistic and n = 2 in diffusive limits, where $N_\uparrow(E_f)$ and $N_\downarrow(E_f)$ are the density of states and $v_{f\uparrow}$ and $v_{f\downarrow}$ are the Fermi velocities for majority and minority spins respectively. To determine the spin polarization values the conductance curves are analyzed using the modified BTK model.[14]

The PCAR measurements in semiconductors (Sm) are often somewhat restricted due to the Schottky barrier present at the Sm/superconductor (S) interface. While by increasing the doping concentrations in semiconductors this problem can be circumvented, as has been demonstrated in Ref. 11, there have been very few experimental studies involving Andreev reflection in semiconductors[15] or dilute magnetic semiconductors,[16,17] compared to more conventional ferromagnetic materials.[2] The situation is further complicated by the fact that a comprehensive microscopic theory of the DMS materials must incorporate both Mn disorder and spin-orbit interaction, and has yet to be completed. Interestingly, in contrast to (In,Mn)Sb[11] evaluating the spin polarization by PCAR for (Ga,Mn)As has not been straightforward[17] as additional pair-breaking mechanism, likely to be due to inelastic quasiparticle scattering[18] had to be taken into account; very similar to the results recently obtained in lithographically defined nanocontacts with the intentionally introduced Pt layer.[19] While the situation in (Ga,Mn)As can be further complicated by the unconventional transport mechanism in these alloys, which is likely to involve impurity band scattering,[17] these results suggest that there may be a correlation between the carrier mobility in a compound and inelastic scattering rate. (Ga,Mn)Sb is a good candidate to further test this hypothesis, as its carrier mobility is intermediate between (In,Mn)Sb and (Ga,Mn)As.

In order to make point contacts with (Ga,Mn)Sb mechanically polished sharp tips of Pb and Sn, with the superconducting transition temperatures $T_c \sim$ 7.2 K and $\sim$ 3.7 K respectively, were used. The conductance curves were measured by the standard lock-in-detection method at 2 kHz. The details of the experimental technique can be found in Ref. 20. The experimental curves were then fitted using the modified BTK model[14] to extract the magnitude of spin polarization $P$ and the interfacial barrier strength, $Z$. The spreading resistance of the film,[21] determined to be $\sim$5 $\Omega$ by the four probe method, was included in the model in addition to the temperature $T$ and the superconducting gap, $\Delta$. The (Ga,Mn)Sb films were grown on GaAs at a substrate temperature $\sim$210$^o$C with a 3 µm ZnTe buffer layer deposited at 300$^o$C by molecular beam epitaxy. The Mn concentration was $\sim$ 2%, the film thickness $\sim$ 300 nm, and the carrier concentration, n $\sim$ 3 $\times$ 10$^{20}$/cm$^3$, similar to (In,Mn)Sb.[11] Further details on the growth technique can be found in Ref. 9.

The magnetization curves of (Ga,Mn)Sb film at different temperatures measured by a SQUID magnetometer are shown in Fig.1. The hysteresis loops suggest ferromagnetic ordering in the film below approximately 10 K. The remnant magnetization curve measured for the magnetic field of ~10 Oe as a function of temperature, shown in the inset of Fig.1, reveals that the Curie temperature, $T_C$ is ~10 K. The presence of small magnetization at temperatures above $T_C$ is likely to indicate the presence of some MnSb precipitates.[22] The temperature dependence of the film resistance displays a corresponding maximum around $T_C$ (see Fig.2). Such maxima have been observed in magnetic metals,[23] magnetic semiconductors,[24] and recently in dilute magnetic semiconductors,[25,26] and is believed to be the result of the critical carrier scattering due to the spin order-disorder transition.

A representative conductance curve obtained with a Sn tip is shown in Fig.3. In order to account for inelastic broadening which is likely to take place in the Sn/(Ga,Mn)Sb contact area we used higher effective temperatures in the fitting routine, which is equivalent to introducing some inelastic quasiparticle broadening $\Gamma$.[18,27] This approach, however, leads to the increase in the number of free parameters, and as was noted in Ref. 27, one has to be careful to make sure that the fitting procedure in this case still remains stable with respect to small perturbations. To test this we have gradually reduced the superconducting gap of Sn from the bulk value of to 0.57 meV to 0.5 meV, which, as can be seen in Fig. 3b resulted in the change in $P$ of only 1%. Fig. 4 shows the normalized conductance curves for a Pb tip contact at different temperatures for a single contact between the superconducting Pb tip and the (Ga,Mn)Sb film. The low temperature curves are fitted using a superconducting gap, $\Delta \sim 1$ meV, which is smaller than the bulk BCS gap of Pb ~ 1.3 meV, possibly due to strong electron-phonon coupling. Similarly to the Sn contacts the effective temperatures used in the fits are higher than the physical temperatures. Importantly, the effective temperatures used to fit both sets of data are very similar and have the lower limit of ~ 4.0 K for all the data with T< 3.5 K, which seem to indicate that the main broadening mechanism in this temperature range is inelastic broadening originating from (Ga,Mn)Sb. The average magnitude of spin polarization in (Ga,Mn)Sb using both Sn and Pb was determined to be 57%±5%.

Using the free electron approximation, we estimated the Fermi velocity, $v_f = \frac{\hbar}{m^*}(3\pi^2 n)^{1/3}$, of light ($m^* \sim 0.06 m_e$) and heavy holes ($m^* \sim 0.3 m_e$) to be ~ $1.8 \times 10^6 \, m/s$ and $7.8 \times 10^5 \, m/s$ respectively. The carrier concentration, $n$, for light and heavy holes was found ~ $2.75 \times 10^{20}/cm^3$ and $0.25 \times 10^{20}/cm^3$, where we used $n \sim n_{lh} + n_{hh}$; $n_{lh}/n_{hh} \sim (m_{lh}^*/m_{lh}^*)^{3/2}$. We can further estimate the minimum Z values for both light and heavy holes based on the Fermi velocity mismatch between superconducting Pb, for example, and (Ga,Mn)Sb using $Z_{min} = (r-1)/2\sqrt{r}$ where $r = v_{Pb}/v_{GaSb}$.[28] The calculated values are $Z_{lh} \sim 0.02$ and $Z_{hh} \sim 0.44$. The extracted Z values have never exceeded 0.02, which, within the accuracy of the model, corresponds to a clean interface, indicating the light-hole only contribution to Z.[29] Using the Drude model, we have also estimated the spin relaxation times,[30] $\tau = m^*/n\rho e^2$, where $\rho_{2K} \sim$ 0.4 mΩ.cm (see Fig. 2); $\tau_{lh} \sim 2.1 \times 10^{-14}$ s and $\tau_{lh} \sim 1 \times 10^{-14}$ s yielding the mean free paths, $l = v_f \tau$, of $l_{lh} \sim 38$ nm and $l_{lh} \sim 8$ nm. The contact size, $d$, is determined from Wexler's expression[31]: $R_c \sim \frac{4}{3\pi}\frac{\rho_o l}{d^2} + \frac{\rho_o}{2d}$. For $R_c \sim 75\ \Omega$, the contact size $\sim 38\ nm$, so the transport corresponds to the ballistic regime for light holes and the intermediate regime for heavy holes ($l_{hh} \sim d$), which is different from the case of (In,Mn)Sb[11], where the transport regime is ballistic for all types of carriers. It is thus reasonable to assume that in order to describe (Ga,Mn)Sb, as well as (Ga,Mn)As,[17] where the transport is purely diffusive, the modified BTK model[14] has to be extended to include inelastic broadening parameter $\Gamma$.[19] $\Gamma_{GaSb}$ can be estimated to be approximately 0.56 meV ($T_{eff} \sim 4$ K) and $\Gamma_{GaAs} \sim 0.77$ meV ($T_{eff} \sim 5.5$ K), where $\Gamma = \sqrt{2}T_{eff}$. The ratio of the two $\Gamma_{GaAs}/\Gamma_{GaSb} \sim 1.4$ can be compared to the ratio of the respective band gaps in (Ga,Mn)As and (Ga,Mn)Sb: $E_{GaAs}/E_{GaSb} = 1.52\ eV/0.82\ eV = 1.85$. While this is a fairly crude estimate, there seems a clear correlation between the degree of inelastic broadening in the S/Sm contacts and the band gap, particularly in view of the fact that (In,Mn)Sb contacts[11] with the smallest band gap of only 0.24 eV does not exhibit any noticeable DOS broadening. Importantly, while for GaMnAs, $P$ is very sensitive to the effective temperature, yielding a large uncertainty of ±17%,[17] for (Ga,Mn)Sb the uncertainly is just ±5%, only slightly inferior to the typical accuracy of the PCAR technique.

In summary, we have measured the transport spin polarization of (Ga,Mn)Sb epitaxially grown films using the PCAR technique. The results for two different superconductors, Sn and Pb, are quite similar, yielding the average $P = 57\pm5\%$, a fairly respectable value. The data is analyzed by introducing a generic quasiparticle spectrum broadening, which is likely to be related to inelastic scattering in the S/Sm contacts, as it has recently been reported in the case of Co-Pt-Pb nanocontact,[19] and as has been argued earlier in the case of (Ga,Mn)As,[17] where this effect is even more pronounced due to the larger band gap and lower carrier mobility.

The authors thank I.I. Mazin for useful discussions. The work at WSU was supported by DARPA through ONR Grant N00014-02-1-0886 and NSF CAREER 0239058 (B.N.) and at Notre-Dame by NSF Grant DMR06-0375 (J.K.F.)

FIGURE CAPTIONS:

Fig.1. Magnetic hysteresis curves of (Ga,Mn)Sb film measured at different temperatures. The applied field, $H$, is parallel to the plane of the film. Inset shows in-plane magnetization as a function of temperature obtained at 10 Oe field. Both curves indicate the Curie temperature of ~ 10 K.

Fig.2. Low temperature resistance of the 300 nm thick (Ga,Mn)Sb film. The maximum at ~9 K is close to the magnetic transition; the estimated resistivity at 2 K ~ 0.4 mΩ·cm.

Fig.3. Typical normalized conductance curves shown for Sn with a contact resistance of $R_c$ = 90 Ω at $T$= 1.2 K. Solid lines are the numerical fits. The effective temperatures are $T_{eff}$ = 4.5 K; a) fit with the bulk BCS gap for Sn $\Delta$ = 0.57 meV. b) fit with the same inelastic broadening but a reduced gap $\Delta$ = 0.5 meV.

Fig.4. Normalized conductance curves for a Pb point contact with $R_c$ = 75 Ω at different temperatures. The solid lines are the numerical fits. Fitting parameters: $\Delta$ = 1.0 meV, Z = 0.003 and the average $P$ ~ 57%. The effective temperatures are $T_{eff}$ ~ 4-4.5 K. Inset shows the extracted spin polarization values at different temperatures.

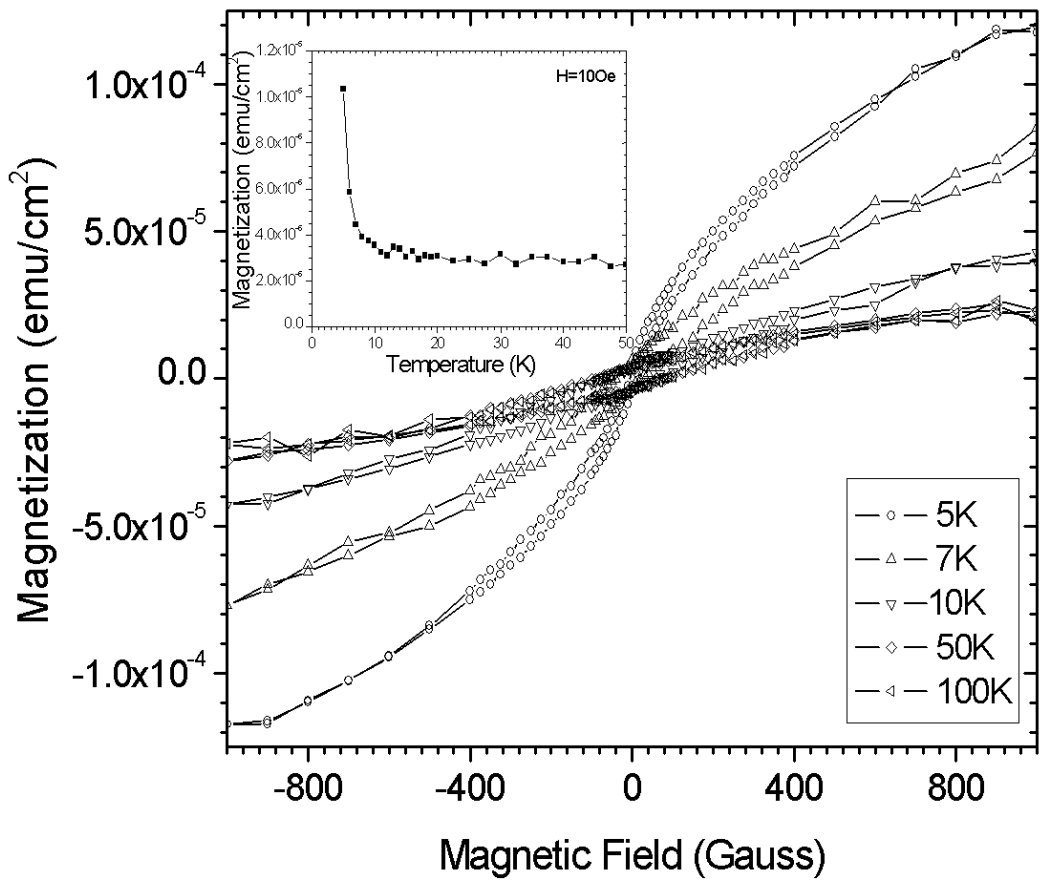

Fig.1

Panguluri, et al.

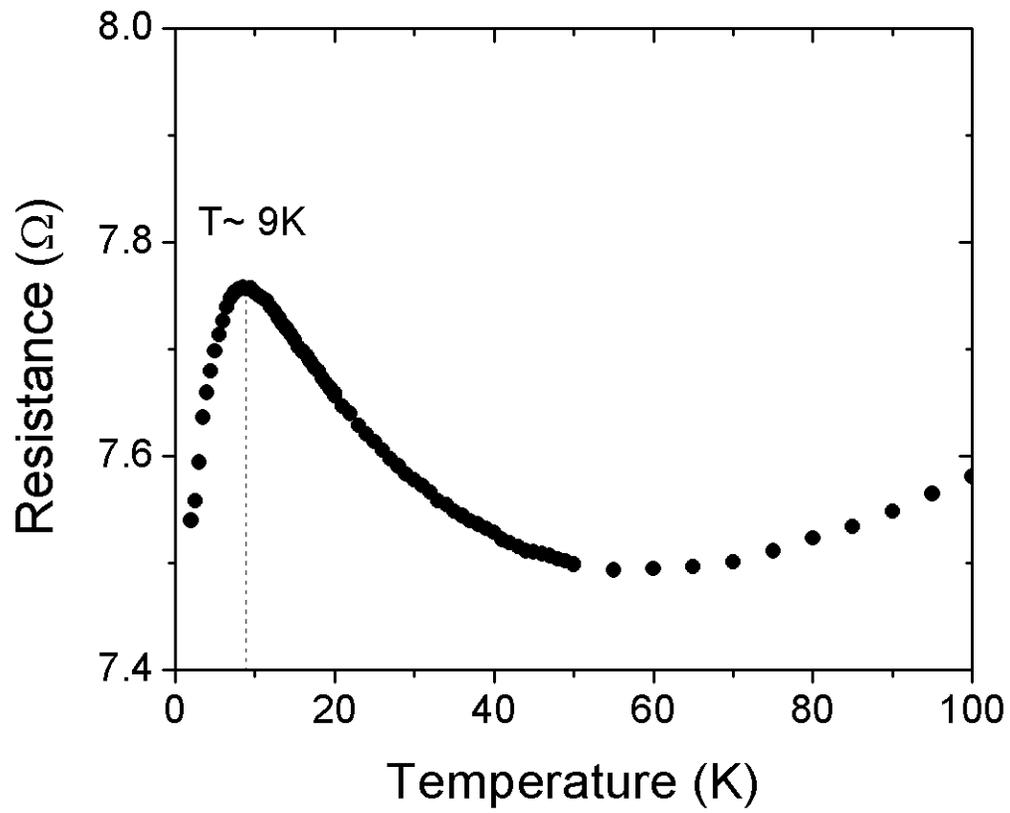

Fig.2

Panguluri, et al.

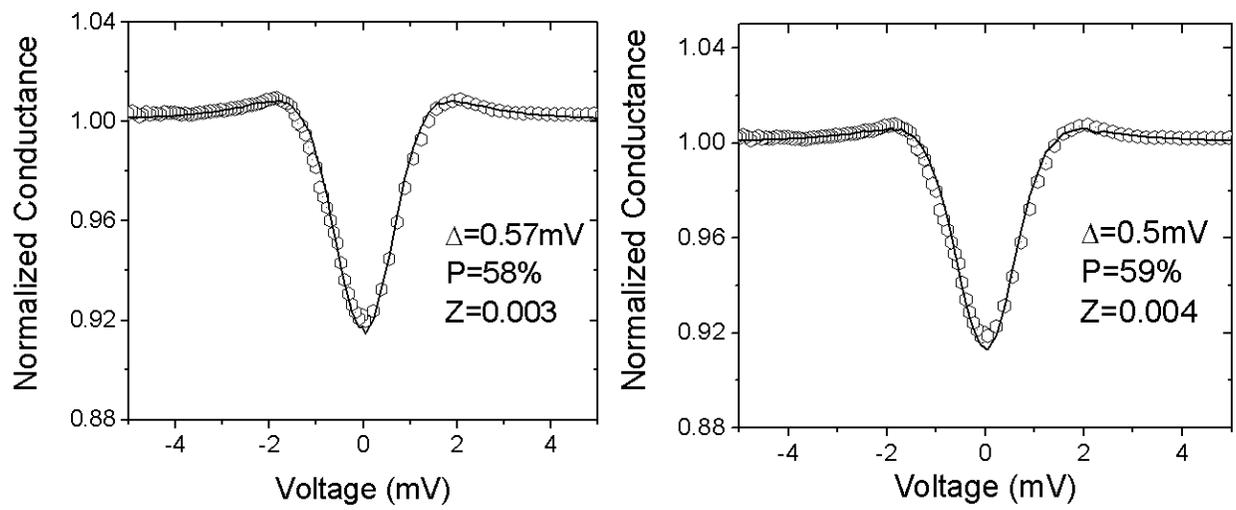

Fig.3

Panguluri, *et. al*.

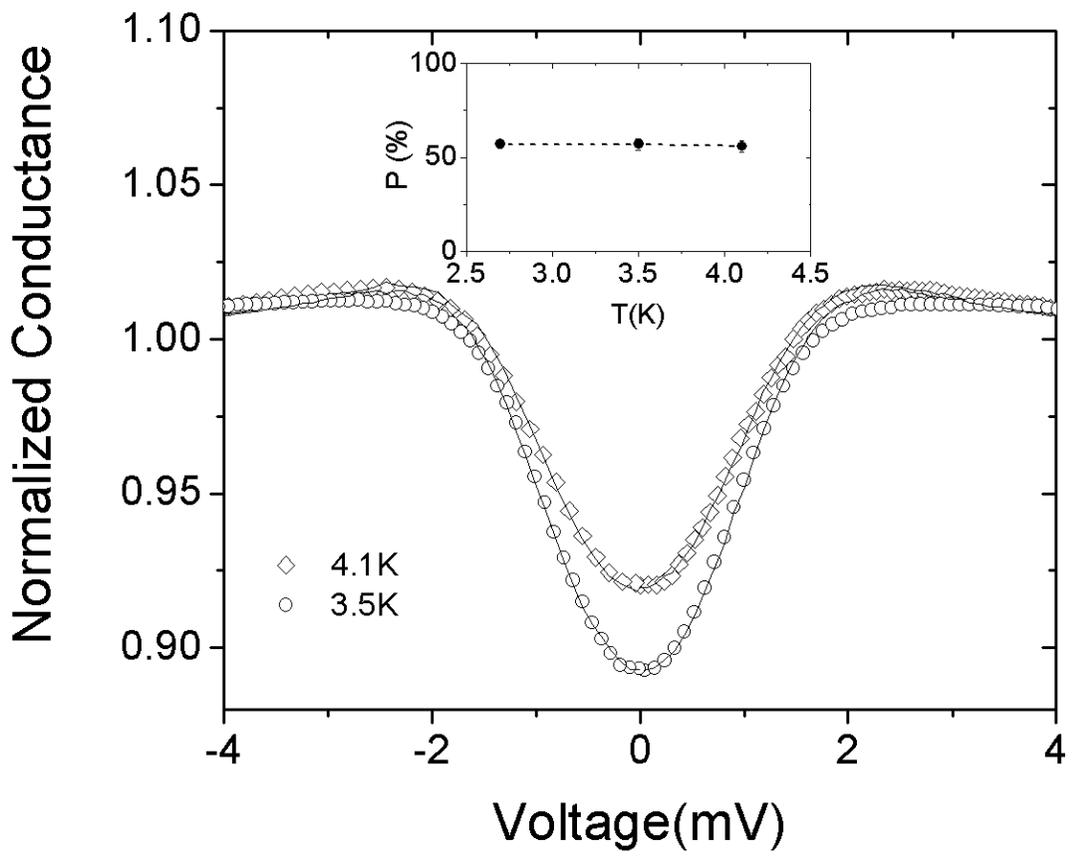

Fig.4

Panguluri *et. al.*